\documentclass[fleqn,11pt]{article}

\usepackage{amsfonts,amsmath,amssymb}
\usepackage{latexsym}

\usepackage{amsfonts,amssymb,cite}
\usepackage{graphicx}

\def\noi{\noindent}
\def\jnumber#1#2{\thispagestyle{empty} \noi\unitlength=1mm
    	\begin{picture}(178,10)
            \put(177,15){\llap{\large\it Grav. Cosmol. No.\,#1, #2}}
                    \end{picture}}

\newcommand{\Title}[1]{\noi {{\Large\bf #1}}\\[1ex]}

\newcommand{\Author}[2]{\noi{\bf #1}\\[2ex]\noi{\normalsize\it #2}\\}

\def\au#1{${}^{#1}$}

\newcommand{\Abstract}[1]{\vskip 2mm \begin{center}
        \parbox{16.4cm}{\small\noi #1} \end{center}\medskip}

\def\email#1#2{\footnotetext[#1]{e-mail: #2}\addtocounter{footnote}{1}}

\def\nqq{\hspace*{-2em}}

\def\cm{\hspace*{1cm}}

\usepackage{color}

\def\Jl#1#2{#1 {\bf #2},\ }

\def\ApJ#1 {\Jl{Astroph. J.}{#1}}
\def\CQG#1 {\Jl{Class. Quantum Grav.}{#1}}
\def\DAN#1 {\Jl{Dokl. AN SSSR}{#1}}
\def\GC#1 {\Jl{Grav. Cosmol.}{#1}}
\def\GRG#1 {\Jl{Gen. Rel. Grav.}{#1}}
\def\IJMPD#1 {\Jl{Int. J. Mod. Phys. D}{#1}}
\def\JETF#1 {\Jl{Zh. Eksp. Teor. Fiz.}{#1}}
\def\JETP#1 {\Jl{Sov. Phys. JETP}{#1}}
\def\JHEP#1 {\Jl{JHEP}{#1}}
\def\JMP#1 {\Jl{J. Math. Phys.}{#1}}
\def\NPB#1 {\Jl{Nucl. Phys. B}{#1}}
\def\NP#1 {\Jl{Nucl. Phys.}{#1}}
\def\PLA#1 {\Jl{Phys. Lett. A}{#1}}
\def\PLB#1 {\Jl{Phys. Lett. B}{#1}}
\def\PRD#1 {\Jl{Phys. Rev. D}{#1}}
\def\PRL#1 {\Jl{Phys. Rev. Lett.}{#1}}

\def\lal{&&\nqq {}}

\def\beq{\begin{equation}}
\def\eeq{\end{equation}}
\def\bear{\begin{eqnarray}}
\def\bearr{\begin{eqnarray} \lal}
\def\ear{\end{eqnarray}}
\def\earn{\nonumber \end{eqnarray}}

\def\yy{\\[5pt] {}}

\begin{document}
\twocolumn[
\jnumber{3}{2023}

\Title{Topological Effects With Inverse Quadratic Yukawa Plus Inverse Square Potential on Eigenvalue Solutions\yy}
	   		
\Author{FAIZUDDIN AHMED\au{1}}
 	     {Department of Physics, University of Science \& Technology Meghalaya, Ri-Bhoi, Meghalaya-793101, India }  	   



 	     

\Abstract{

In this analysis, we study the non-relativistic Schrodinger wave equation under the influence of quantum flux field with interactions potential in the background of a point-like global monopole (PGM). In fact, we consider an inverse quadratic Yukawa plus inverse square potential and derive the radial equation employing the Greene-Aldrich approximation scheme in the centrifugal term. We determine the approximate eigenvalue solution using the parametric Nikiforov-Uvarov method and analyze the result. Afterwards, we derive the radial wave equation using the same potential employing a power series expansion method in the exponential potential and solve it analytically. We show that the energy eigenvalues are shifted by the topological defects of a point-like global monopole compared to the flat space result. In addition, we see that the energy eigenvalues depend on the quantum flux field that shows an analogue to the Aharonov-Bohm effect.

}
\medskip

] 
\email 1 {\bf faizuddinahmed15@gmail.com; faizuddin@ustm.ac.in\\ \cm}

\section{Introduction}

The exact and approximate eigenvalue solutions of Schrodinger equation (SE) for some physical potentials, such as the Kratzer potential, pseudoharmonic potential, the Hellmann potential, the Yukawa potential etc. are of highly important in different branches of physics and chemistry \cite{SMI,SMI2,SMI3,CB,CB2,SMI4,SMI5,SMI6,HH2,SHD,BJF} since these solutions contain all the information for a quantum system under investigation. In literature, hydrogen atom and harmonic oscillator problems are given in many textbooks as these two are the only exactly solvable problems quantum system \cite{WG,LDL,HH,FC,DJG}.

The study of quantum mechanical problems under the effects of topological defect is of current research interest in recent times. These effects include the cosmic strings, global monopoles, space-time with a spacelike dislocation and space-time with distortion of a vertical line into a vertical spiral. The conical singularity via a point-like global monopole has been studied, for example, in the non-relativistic limit, the harmonic oscillator problem \cite{CF} and with a potential in \cite{RV}, scattering of charged particles \cite{ALCO2,ERBM33,ERBM44}, quantum motions of spin-zero scalar particles under a magnetic flux with scalar potential \cite{ALCO}, and non-relativistic particle interacts with potential, such as Kratzer potential and Morse potential \cite{VBB}, generalized Morse potential \cite{PN}, diatomic molecular potential \cite{MPP}, generalized Cornell potential (harmonic oscillator plus Mie-type) potential \cite{MPP2},  pseudoharmonic- and Mie-type potential \cite{MPP3} and some other in Refs. \cite{MPP4,MPP5,MPP6,MPP7,MPP8}. A few other investigations of topological defects space-time in the context of non-relativistic quantum systems are in \cite{WCFS,WCFS2}.  However, a point-like global monopole with inverse square plus inverse quadratic Yukawa potential (IQYP) in the presence of the quantum flux hasn't yet been studied which is our aim in this contribution.

Therefore, a static and spherically symmetric space-time describing a point-like global monopole in the spherical coordinates $(r, \theta, \phi)$ in the context of Einstein's general relativity is given by \cite{CF,RV,ALCO2,ERBM33,ERBM44,ALCO,VBB,PN,MPP,MPP2,MPP3,MPP4,MPP5,MPP6,MPP7,MPP8}
\begin{equation}
ds^2_{3D}=g_{ij}\,dx^{i}\,dx^{j},\quad i,j=1,2,3,
\label{1}
\end{equation}
where the metric tensor $g_{ij}$ and its inverse are given by
\begin{equation}
g_{ij}=\begin{pmatrix}
\frac{1}{\alpha^2} & 0 & 0\\
0 & r^2 & 0\\
0 & 0 & r^2\,\sin^2 \theta
\end{pmatrix},\, 
g^{ij}=\begin{pmatrix}
\alpha^2 & 0 & 0\\
0 & \frac{1}{r^2} & 0\\
0 & 0 & \frac{1}{r^2\,\sin^2 \theta}
\end{pmatrix}.
\label{2}
\end{equation}
Here $0 < \alpha \leq 1$ characterise the topological defect parameter of point-like global monopole (PGM). One of the interesting feature of this geometry in four-dimension is that this metric possesses a curvature singularity on the axis $R=\frac{2\,(1-\alpha^2)}{r^2}$. Other properties of this conical singularity space-time were given in details in Refs. \cite{CF,ALCO2,ERBM33,ERBM44,ALCO}. For $\alpha \to 1$, the space-time reduces to Minkowski flat space $ds^2_{3D}=dr^2+r^2\,d\Omega^2$, where $d\Omega^2=\left(d\theta^2+\sin^2 \theta\,d\phi^2\right)$. It is worth mentioning that the presence of topological defect via a point-like global monopole or cosmic string space-time modify the eigenvalue solution and shifts the result compared to flat space case. However, the quantum systems in the non-relativistic limit in the background of some topological defect geometries other than point-like global monopole and cosmic string give rise to an analogue of the Aharonov-Bohm-type effect (see, Refs. \cite{WCFS,WCFS2} and related references).

This paper is organised as follows: in {\tt section 2}, we discuss the Schr\"{o}dinger wave equation with a potential under the influence of the quantum flux field in the background of a point-like global monopole. We derive the radial equation with inverse quadratic Yukawa plus centrifugal potential and later on with an inverse quadratic Yukawa potential employing the Greene-Aldrich approximate scheme and solve the equations; in {\tt section 3}, we derive the radial equation of the same quantum system employing a power series expansion and obtain the eigenvalue solution; in {\tt section 4}, we present our conclusions. We have used the natural units $c=1=\hbar$.

\section{Schr\"{o}dinger Particles in Point-like Global Monopole with Potential: Greene-Aldrich Approximation Scheme }
\label{sec: 1}

In this section, we determine the eigenvalue solution of non-relativistic particles under the influence of the quantum flux field in the background of a point-like global monopole with potential. We use the Greene-Aldrich approximate scheme and solve the radial equation using the parametric NU-method and discuss the effects of factors, such as the topological defect and the magnetic flux. We see that the topological defect of point-like global monopole shifts the energy levels compared to the flat space results.

The time-dependent Schr\"{o}dinger wave equation is described by the following equation \cite{CF,RV,VBB,PN,MPP,MPP2,MPP3,MPP4,MPP5,MPP6,MPP7,MPP8,WCFS,WCFS2}
\begin{eqnarray}
\Big[-\frac{1}{2\mu}\frac{1}{\sqrt{g}}\,D_{i}\,\Big(\sqrt{g}\,g^{ij}\,D_{j}\Big)+V(r)\Big]\,\Psi=i\,\frac{\partial\,\Psi}{\partial\,t},
\label{3}
\end{eqnarray}
where $D_{i} \equiv \Big(\partial_{i}-i\,e\,A_{i}\Big)$ \cite{WG,LDL}, $i=1,2,3$ with $e$ is the electric charges, $\vec{A}$ is the electromagnetic three-vector potential, $g=|g_{ij}|$ is the determinant of the metric tensor $g_{ij}$ with $g^{ij}$ its inverse. In this analysis, we have chosen the electromagnetic three-vector potential $\vec{A}$ given by $A_{r}=0=A_{\theta},\quad A_{\phi}=\frac{\Phi_{AB}}{2\,\pi\,r\,\sin \theta}, \Phi_{AB}=\Phi\,\Phi_0,\Phi_0=2\,\pi\,e^{-1},$ Refs. \cite{ALCO,ABHA,MPP,MPP2,MPP3,MPP4,MPP5,MPP6,MPP7,MPP8} where, $\Phi_{AB}=const$ is the Aharonov-Bohm magnetic flux, $\Phi_0$ is the quantum of magnetic flux, and $\Phi$ is the amount of magnetic flux which is a positive integer. 

In quantum mechanical problems, the total wave function $\Psi (t, r, \theta, \phi)$ in terms of a radial wave function $\psi (r)$ can express as $\Psi(t, r, \theta, \phi)=e^{-i\,E\,t}\,Y_{l,m} (\theta, \phi)\,\frac{\psi (r)}{r}$ where, $E$ is the energy of the Schrodinger particle, $Y_{l,m} (\theta, \phi)=A_{l,m} (\theta)\,B_{m} (\phi)$ is the spherical harmonic functions, and $l, m$ are respectively the angular momentum and magnetic moment quantum numbers.

Thereby, in the space-time background (\ref{1}) and using the electromagnetic potential and total wave function in the Eq. (\ref{3}), we obtain the following differential equation:
\begin{eqnarray}
\psi''(r)+\frac{1}{\alpha^2}\,\Bigg[2\,\mu\,\Big(E-V(r)\Big)-\frac{l'\,(l'+1)}{r^2}\Bigg]\,\psi (r)=0,
\label{6}
\end{eqnarray}
where $l'=(l-\Phi)$. 


\subsection{\bf Interactions with Inverse Quadratic Yukawa Plus Inverse Square Potential}


In this section, we study the quantum motions of the non-relativistic particles interact with a spherically symmetric potential $V(r)$ in the background of a point-like global monopole under the influence of the quantum flux field. This potential is given by
\begin{equation}
V (r)=\frac{1}{r^2}\,\Big(V_1-V_2\,e^{-2\,\delta\,r}\Big),
\label{8}
\end{equation}
where $V_1, V_2$ characterises the strength of potentials, and $\delta$ is the screening parameter. Noted that for $V_1 \to 0$, the potential (\ref{8}) reduces to inverse quadratic Yukawa potential (IQYP) given by $V (r)=-\frac{V_2}{r^2}\,e^{-2\,\delta\,r}$ \cite{MH2,CAO,BII2}. While for $V_2 \to 0$, we have inverse square potential $V (r) \sim \frac{1}{r^2}$. 

\begin{figure*}
\centering
\includegraphics[width=2.9in,height=1.6in]{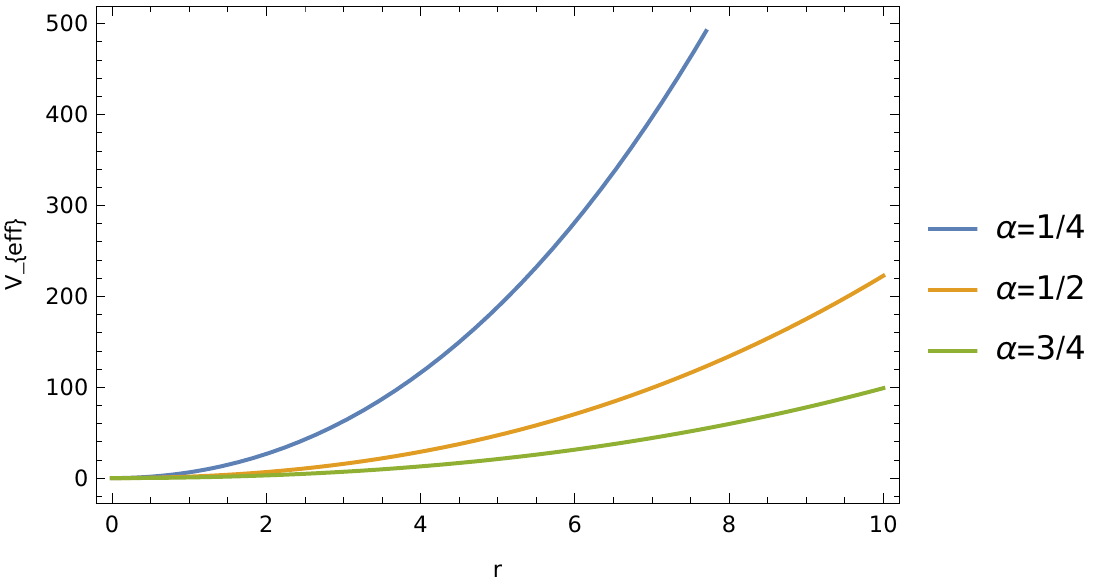}\quad\quad\quad
\includegraphics[width=2.9in,height=1.6in]{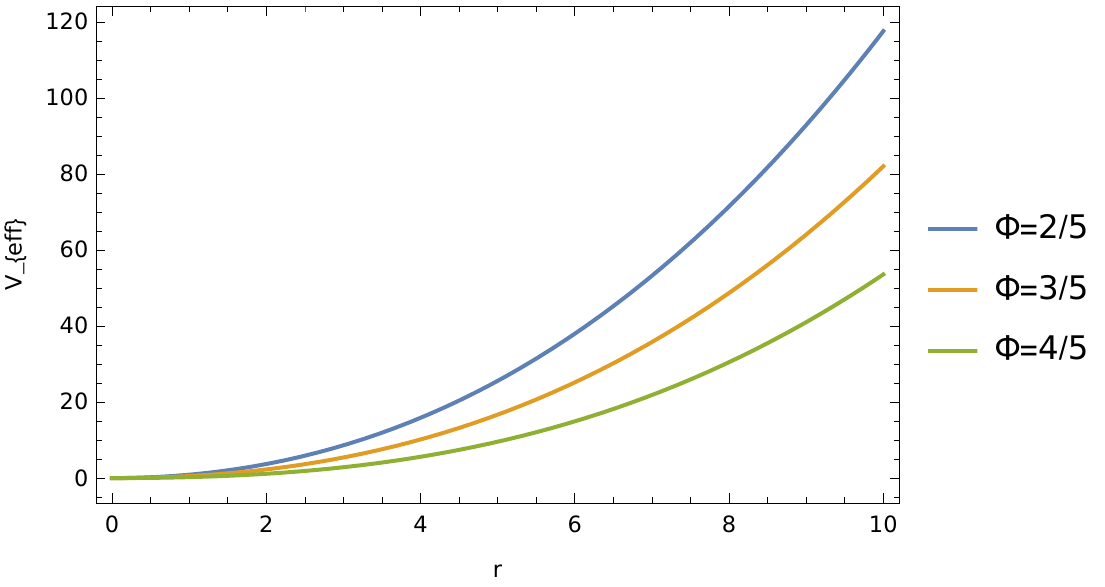}\quad
\includegraphics[width=3.2in,height=1.6in]{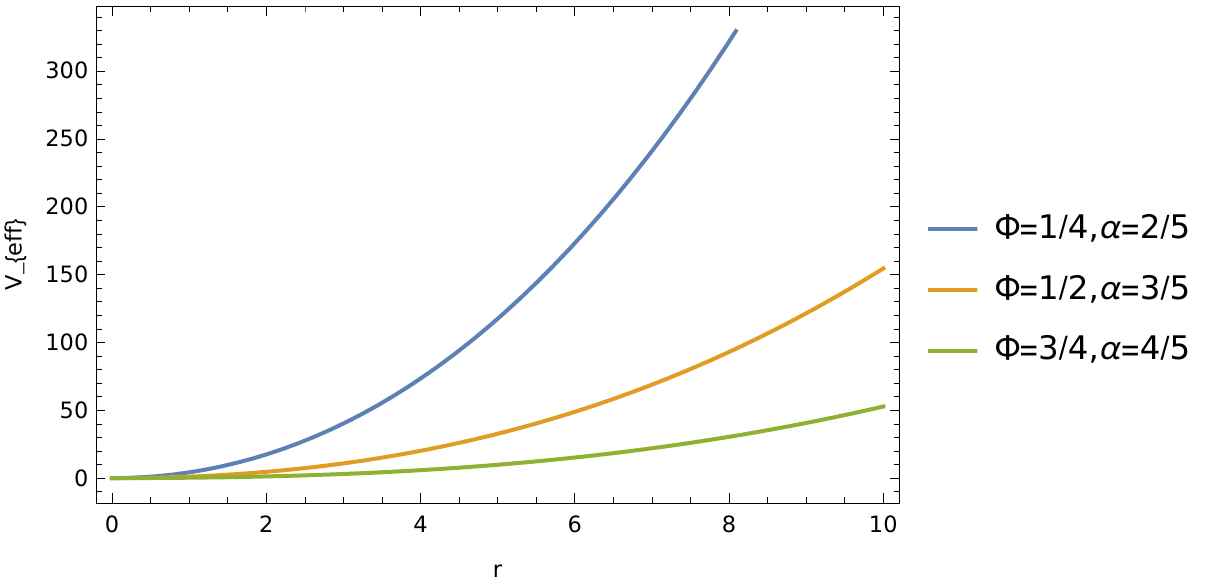}\quad
\caption{\small Effective potential $V_{eff}$ with radial distance $r$. Top left: different values of the topological defect parameter $\alpha$ keeping fixed $\Phi=1/2$; Top right: different values of quantum flux keeping fixed the topological parameter $\alpha=3/4$; and bottom: different values of the topological defect parameter and quantum flux. Here, we have set the parameters $l=1=M=V_1=V_2$, $\delta=0.01$ and the units of these parameters are chosen in a system of units where $c=1=\hbar=G$.}
\label{fig: 1}
\end{figure*}

Using the above potential (\ref{8}), one can easily obtain the effective potential of the quantum system in a curved space-time background as follows
\begin{eqnarray}
V_{eff} (r)=\frac{1}{2\,\mu\,\alpha^2\,r^2}\,\Big[(l-\Phi)\,(l-\Phi+1)\nonumber\\
+2\,\mu\,(V_1-V_2\,e^{-2\,\delta\,r})\Big].
\label{sp}
\end{eqnarray}
One can see that the effective potential of the quantum system in a curved space-time background depends on the topological defect of the geometry characterized by the parameter $\alpha$ and the magnetic flux. For $\alpha \to 1$, the geometry under consideration becomes Minkowski flat space. We have plotted a few graphs showing the effects of various parameters on the effective potential (fig. 1). 

Thereby, substituting the potential (\ref{8}) in the Eq. (\ref{6}), we obtain the following radial wave equation:
\begin{equation}
\psi'' (r)+\Bigg[\beta-\frac{\tau^2}{r^2}-\frac{2\,\mu}{\alpha^2\,r^2}\,\Big(V_1-V_2\,e^{-2\,\delta\,r}\Big)\Bigg]\,\psi (r)=0,
\label{9}
\end{equation}
where 
\begin{equation}
\beta=\frac{2\,\mu\,E}{\alpha^2}\quad,\quad \tau=\sqrt{\frac{(l-\Phi)\,(l-\Phi+1)}{\alpha^2}}.
\label{10}
\end{equation}

The radial part of the Schr\"{o}dinger equation for this potential can be solved exactly for $l=0$
(s-wave) but cannot be solved for $l \neq 0$ easily. To obtain the solution for $l \neq 0$, we must employ a suitable approximation scheme and we have considered the Greene-Aldrich approximation scheme \cite{RLG} to deal with the centrifugal term, which is given as follows:
\begin{equation}
\frac{1}{r^2} \approx \frac{4\,\delta^2}{(1-e^{-2\,\delta\,r})^2}.
\label{11}
\end{equation}
It is noted that for a short-range potential, the relation (\ref{11}) is a good approximation to $1/r^2$ \cite{RLG,MRS,WCQ}. This approximation is valid for small values of the screening parameter $\delta << 1$.

Therefore, employing the above approximation in Eq. (\ref{10}), we obtain  
\begin{equation}
\psi ''(r)+\Bigg[\beta-\frac{\Big(\beta_1-\beta_2\,e^{-2\,\delta\,r}\Big)}{(1-e^{-2\,\delta\,r})^2}\Bigg]\,\psi (r)=0,
\label{12}
\end{equation}
where we set the parameters
\begin{eqnarray}
\beta_1=4\,\delta^2\,\Big(\tau^2+\frac{2\,\mu\,V_1}{\alpha^2}\Big),\quad \beta_2=\frac{8\,\mu\,V_2\,\delta^2}{\alpha^2}.
\label{13}
\end{eqnarray}

Let us perform a change of new variable via $s=e^{-2\,\delta\,r}$ into the above Eq. (\ref{12}), we obtain the following second-order differential equation:
\begin{equation}
\psi''(s)+\frac{(c_1-\,c_2\,s)}{s\,(1-c_3\,s)}\,\psi'(s)+\frac{(-\xi_1\,s^2+\xi_2\,s-\xi_3)}{s^2\,(1-c_3\,s)^2}\,\psi (s)=0,
\label{14}
\end{equation}
where $c_1=1=c_2=c_3$ and 
\begin{equation}
\xi_1=-\frac{\beta}{4\,\delta^2},\quad \xi_2=-\frac{1}{4\,\delta^2}\,(2\,\beta-\beta_2),\quad \xi_3=\frac{1}{4\,\delta^2}\,(\beta_1-\beta).
\label{15}
\end{equation}
The above differential equation can solve using a well-known method called the the Nikiforov-Uvarov method \cite{AFN}. This method is very much helpful in order to find the eigenvalues and eigenfunction of the Schr\"{o}dinger-like equation, as well as other second-order differential equations of physical interest. Several authors have been successfully applied this method in order to obtain the eigenvalue solutions of the wave equation (see, Refs. \cite{SZ,SZ2,MPP,MPP2,MPP4,MPP5,MPP6,MPP7,MPP8}).

Thus, by comparing Eq. (\ref{14}) with the Eq. (A.1) in appendix, we have the following coefficients
\begin{eqnarray}
&&c_4=0,\quad c_5=-\frac{1}{2},\quad c_6=\frac{1}{4}+\xi_1,\quad c_7=-\xi_2,\nonumber\\
&&c_8=\xi_3,\quad c_9=\frac{1}{4}+\xi_1-\xi_2+\xi_3,\quad c_{10}=1+2\,\sqrt{\xi_3},\nonumber\\
&&c_{11}=2\,\Big(1+\sqrt{\frac{1}{4}+\xi_1-\xi_2+\xi_3}+\sqrt{\xi_3}\Big),\nonumber\\
&&c_{12}=\sqrt{\xi_3},\nonumber\\
&&c_{13}=-\frac{1}{2}-\Big(\sqrt{\frac{1}{4}+\xi_1-\xi_2+\xi_3}+\sqrt{\xi_3}\Big).
\label{20}
\end{eqnarray}

Substituting Eq. (\ref{20}) into the Eq. (A.3) in appendix and using Eq. (\ref{15}), one can obtain the following the energy eigenvalues expression 
\begin{eqnarray}
E_{n,l}&=&\frac{2\,\alpha^2\,\delta^2}{\mu}\,\Bigg[\frac{(l-\Phi)\,(l-\Phi+1)+2\,\mu\,V_1}{\alpha^2}\nonumber\\
&&-\Bigg(\frac{\sum}{2}+\frac{(l-\Phi)\,(l-\Phi+1)+2\,\mu\,V_1}{2\,\sum\,\alpha^2}\Bigg)^2\Bigg],\quad\quad
\label{21}
\end{eqnarray}
where 
\begin{eqnarray}
&&\sum=\Big(n+\frac{1}{2}+\frac{1}{\alpha}\times\nonumber\\
&&\sqrt{\frac{\alpha^2}{4}+(l-\Phi)\,(l-\Phi+1)+2\,\mu\,(V_1-V_2)}\Big).\quad\quad
\label{ee}
\end{eqnarray}

The radial wave functions are given by
\begin{equation}
\psi_{n,l} (s)=s^{\sigma}\,(1-s)^{\sum-n}\,P^{(2\,\sigma\,,\,2\,\sum-2\,n-1)}_{n}\,(1-2\,s),
\label{22}
\end{equation}
where $\sigma=\Bigg(\frac{\sum}{2}+\frac{(l-\Phi)\,(l-\Phi+1)+2\,\mu\,V_1}{2\,\sum\,\alpha^2}\Bigg)$ and $P^{(a,b)}_{n} (x)$ are Jacobi polynomials.

Equation (\ref{21}) is the non-relativistic energy profile and Eq. (\ref{22}) is the radial wave functions of a non-relativistic particle under the influence of the quantum flux with inverse quadratic plus inverse quadratic Yukawa potential in a topological defect space-time produced by a point-like global monopole. We see that the eigenvalue solution gets modified compared to the flat space results due to the presence of the topological defects of the geometry characterized by the parameter $\alpha$ and shifted the results. We also see that the energy levels shifted due to the presence of the magnetic flux and this dependence of the eigenvalue on the geometric quantum phase gives us an analogous of the Aharonov-Bohm effect \cite{YA,MP}.

If we analyze the above quantum system without topological effects, that is, $\alpha \to 1$. In that case, the space-time geometry (\ref{1}) under consideration becomes flat space. Therefore, for $\alpha \to 1$, the energy eigenvalue from Eq. (\ref{21}) becomes
\begin{eqnarray}\nonumber
&&E_{n,l}=\frac{2\,\delta^2}{\mu}\,\Bigg[(l-\Phi)\,(l-\Phi+1)+2\,\mu\,V_1\nonumber\\
&&-\Bigg(\frac{\Pi}{2}+\frac{(l-\Phi)\,(l-\Phi+1)+2\,\mu\,V_1}{2\,\Pi}\Bigg)^2\Bigg],
\label{25}
\end{eqnarray}
where 
\begin{equation}
\Pi=\Big(n+\frac{1}{2}+\sqrt{\Big(l-\Phi+\frac{1}{2}\Big)^2+2\,\mu\,(V_1-V_2)}\Big).
\label{ee2}
\end{equation}
And that the radial wave function from (\ref{22}) becomes 
\begin{equation}
R_{n,l} (s)=s^{\lambda}\,(1-s)^{\Pi-n}\,P^{(2\,\lambda\,,\, 2\,\Pi-2\,n-1)}_{n}\,(1-2\,s),
\label{26}
\end{equation}
where $\lambda=\Bigg(\frac{\Pi}{2}+\frac{(l-\Phi)\,(l-\Phi+1)+2\,\mu\,V_1}{2\,\Pi}\Bigg)$.

Here, only the magnetic flux present in the quantum system shifts the energy levels and the radial function of a non-relativistic particle compared to the flat space results with this superposed potential. 

\begin{figure*}
\centering
\includegraphics[width=2.9in,height=1.7in]{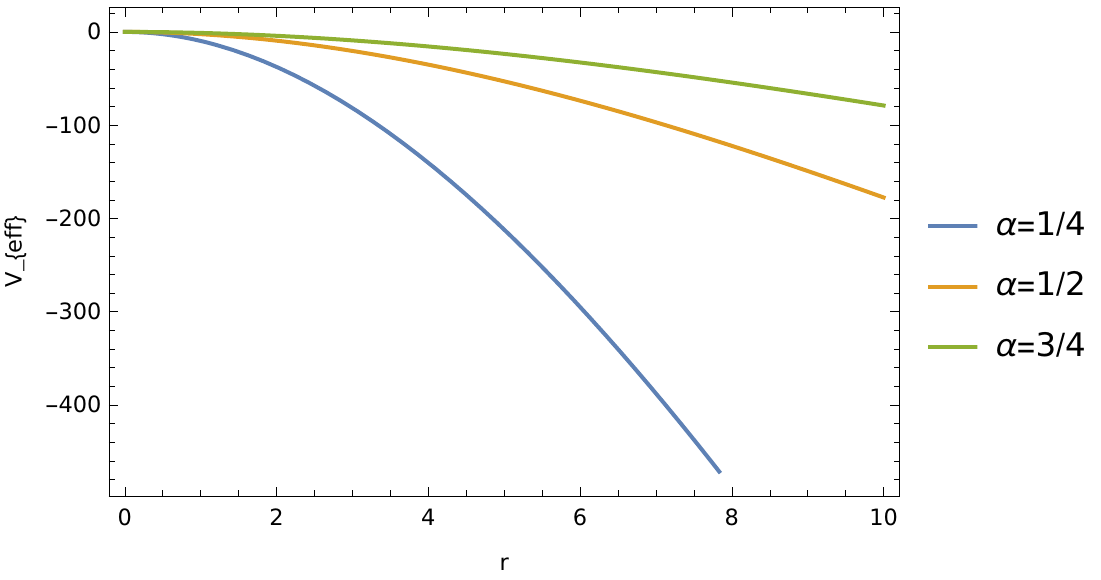}\quad
\includegraphics[width=2.9in,height=1.7in]{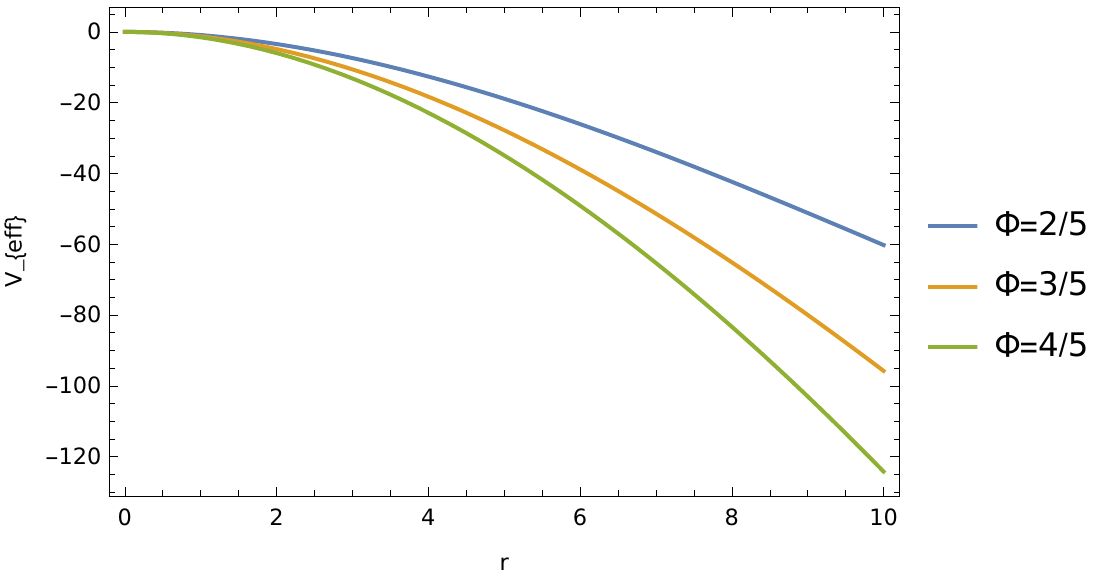}\quad
\includegraphics[width=3.1in,height=1.7in]{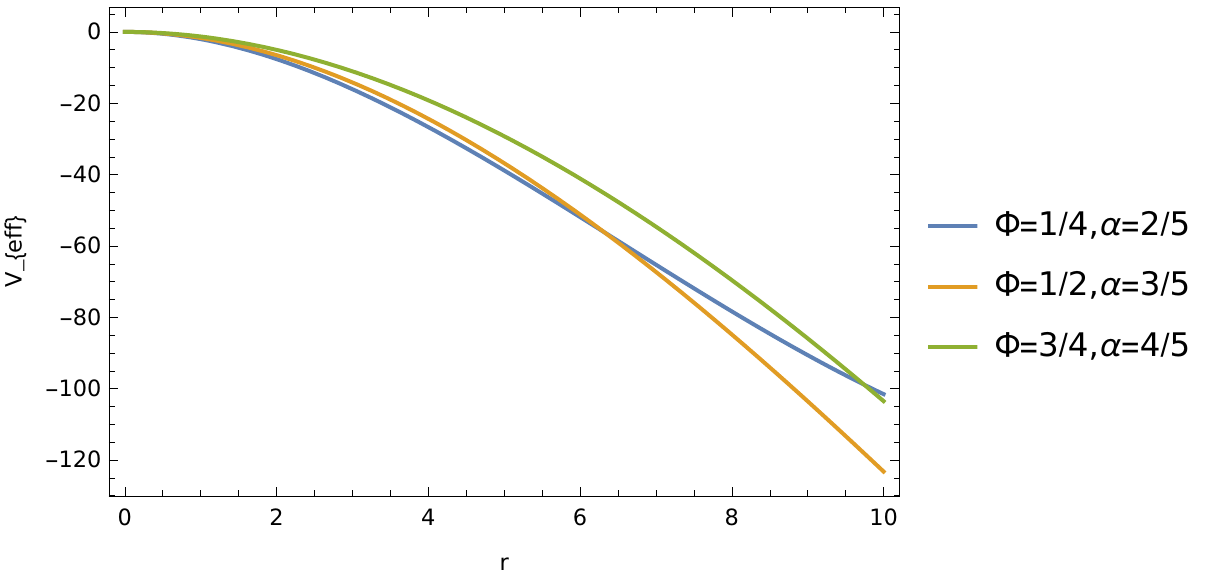}\quad
\caption{\small The effective potential $V_{eff}$ of the system with radial distance $r$. Top Left: with different values of the topological defect parameter $\alpha$ keeping fixed the magnetic flux $\Phi=1/2$; Top right: with different values of the magnetic flux $\Phi$ keeping fixed the topological parameter $\alpha=3/4$; and bottom: with different values of the topological defect parameter $\alpha$ and magnetic flux $\Phi$. Here, we have set the parameters $l=1=M=V_2$, $\delta=0.01$ and the units of these parameters are chosen in a system of units where $c=1=\hbar=G$.}
\label{fig: 2}
\end{figure*}

\subsection{\bf Inverse Quadratic Yukawa Potential}

In this section, are interest on inverse quadratic Yukawa potential (IQYP) given by
\begin{equation}
V (r)=-\frac{V_2}{r^2}\,e^{-2\,\delta\,r},
\label{aa1}
\end{equation}
where $V_2, \delta$ are mentioned earlier. This Yukawa potential was first proposed in \cite{MH2} and later on have been studied the wave equations with this kind of potential in Refs. \cite{CAO,BII2}. 

The effective potential of the quantum system using the inverse quadratic potential becomes
\begin{equation}
V_{eff} (r)=\frac{-2\,\mu\,V_2\,e^{-2\,\delta\,r}+(l-\Phi)\,(l-\Phi+1)}{2\,\mu\,\alpha^2\,r^2}.
\label{eff}
\end{equation}
We have plotted few graphs showing the effective potential for different values of physical parameters involve in it (fig. 2).

Thereby, substituting this IQYP (\ref{aa1}) into the Eq. (\ref{8}), we have 
\begin{equation}
\psi''(r)+\Bigg[\beta-\frac{\tau^2}{r^2}+\frac{2\,\mu\,V_2}{\alpha^2\,r^2}\,e^{-2\,\delta\,r}\Bigg]\,\psi(r)=0,
\label{aa2}
\end{equation}
where $\beta, \tau$ are defined in Eq. (\ref{10}).

Now, using the same approximation (\ref{11}) into the Eq. (\ref{aa2}), we have  
\begin{equation}
\psi''(r)+\Bigg[\beta+\frac{\beta_2\,e^{-2\,\delta\,r}}{(1-e^{-2\,\delta\,r})^2}\Bigg]\,\psi(r)=0,
\label{aa3}
\end{equation}
where $\beta_2$ is given in Eq. (\ref{13}).

Performing a change of variable via $s=e^{-2\,\delta\,r}$ into the above Eq. (\ref{aa3}) and following the same procedure done earlier, one can easily obtain the following expression of the energy eigenvalue
\begin{eqnarray}
&&E_{n,l}=\frac{2\,\alpha^2\,\delta^2}{\mu}\,\Bigg[\frac{(l-\Phi)\,(l-\Phi+1)}{\alpha^2}\nonumber\\
&&-\Bigg(\frac{\Upsilon}{2}+\frac{(l-\Phi)\,(l-\Phi+1)}{2\,\Upsilon\,\alpha^2}\Bigg)^2\Bigg],
\label{aa4}
\end{eqnarray}
where 
\begin{equation}
\Upsilon=\Big(n+\frac{1}{2}+\frac{1}{\alpha}\,\sqrt{\frac{\alpha^2}{4}+(l-\Phi)\,(l-\Phi+1)-2\,\mu\,V_2}\Big).
\label{ee3}
\end{equation}

The radial wave functions are given by  
\begin{equation}
\psi_{n,l} (s)=s^{\kappa}\,(1-s)^{\Upsilon-n}\,P^{(2\,\kappa\,,\,2\,\Upsilon-2\,n-1)}_{n}\,(1-2\,s),
\label{aa5}
\end{equation}
where $\kappa=\Bigg(\frac{\Upsilon}{2}+\frac{(l-\Phi)\,(l-\Phi+1)}{2\,\Upsilon\,\alpha^2}\Bigg)$.

Equation (\ref{aa4}) is the non-relativistic energy profile and Eq. (\ref{aa5}) is the radial wave function in the presence of the Aharonov-Bohm flux field with an inverse quadratic Yukawa potential under the effects of background curvature. We can see that the eigenvalue solution gets modified in comparison the results in Refs. \cite{MH2,CAO,BII2} due to the presence of topological defects defined by the parameter $\alpha$ and the background curvature effects associated with topological defect. We also see that the energy levels shifts due to the presence of magnetic flux and this dependence of the eigenvalue on the geometric quantum phase gives us an analogue of the Aharonov-Bohm effect \cite{YA,MP}.

If one would analyze the quantum system without topological effects, that is, $\alpha \to 1$. In that case, the space-time geometry (\ref{1}) becomes Minkowski flat space. Therefore, for $\alpha \to 1$, the energy eigenvalue from Eq. (\ref{aa4}) becomes
\begin{eqnarray}
&&E_{n,l}=\frac{2\,\delta^2}{\mu}\,\Bigg[(l-\Phi)\,(l-\Phi+1)\nonumber\\
&&-\Bigg(\frac{\Gamma}{2}+\frac{(l-\Phi)\,(l-\Phi+1)}{2\,\Gamma}\Bigg)^2\Bigg],
\label{aa8}
\end{eqnarray}
where 
\begin{equation}
\Gamma=\Big(n+\frac{1}{2}+\sqrt{\Big(l-\Phi+\frac{1}{2}\Big)^2-2\,\mu\,V_2}\Big).
\label{ee4}
\end{equation}

And the radial wave function from Eq. (\ref{aa5}) becomes 
\begin{equation}
R_{n,l} (s)=s^{\zeta}\,(1-s)^{\Gamma-n}\,P^{(2\,\zeta\,,\,2\,\Gamma-2\,n-1)}_{n}\,(1-2\,s),
\label{aa9}
\end{equation}
where $\zeta=\Bigg(\frac{\Gamma}{2}+\frac{(l-\Phi)\,(l-\Phi+1)}{2\,\Gamma}\Bigg)$.

We can see that the magnetic flux present in the quantum system modified the eigenvalue solution (\ref{aa8})--(\ref{aa9}) and shifts the result in flat space background with this inverse quadratic Yukawa potential. . 

\section{Schr\"{o}dinger Particles Confined by AB-flux in Point-like Defect with Potential: Power series Expansion}
\label{sec: 2}

In this section, we study the same quantum system with same inverse quadratic Yukawa plus inverse square potential in the topological defects space-time background. But, we derive here the radial equation by applying a power series expansion in the exponential term that appears in the equation (\ref{9}) and will solve it analytically. 

Expanding the exponential term in potential expression (\ref{8}) up to the second-order, we have
\begin{equation}
\frac{1}{r^2}\,e^{-2\,\delta\,r}=\frac{1}{r^2}-\frac{2\,\delta}{r}+2\,\delta^2.
\label{cc1}
\end{equation}
Therefore, the potential (\ref{8}) can be written as
\begin{equation}
V(r)=\frac{A}{r^2}+\frac{B}{r}-C,
\label{cc2}
\end{equation}
where $A=V_1-V_2$, $B=2\,\delta\,V_2$, and $C=2\,\delta^2\,V_2$.

Therefore, the effective potential of the quantum system 
\begin{eqnarray}
&&V_{eff}=\frac{1}{\alpha^2}\,\Bigg[\frac{(l-\Phi)\,(l-\Phi+1)}{2\,\mu\,r^2}+\frac{V_1-V_2}{r^2}\nonumber\\
&&+\frac{2\,\delta\,V_2}{r}-2\,\delta^2\,V_2\Bigg].
\label{cc}
\end{eqnarray}
We have plotted few graphs showing the effective potential for different values of physical parameters involve in it (fig. 3).

\begin{figure*}
\centering
\includegraphics[width=2.8in,height=1.6in]{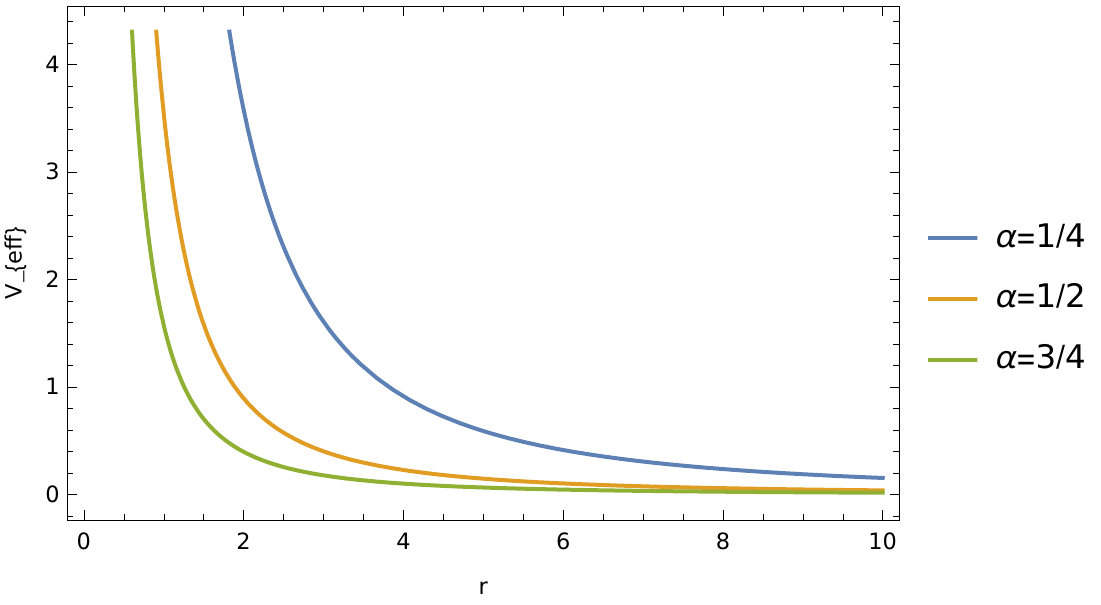}\quad
\includegraphics[width=2.8in,height=1.6in]{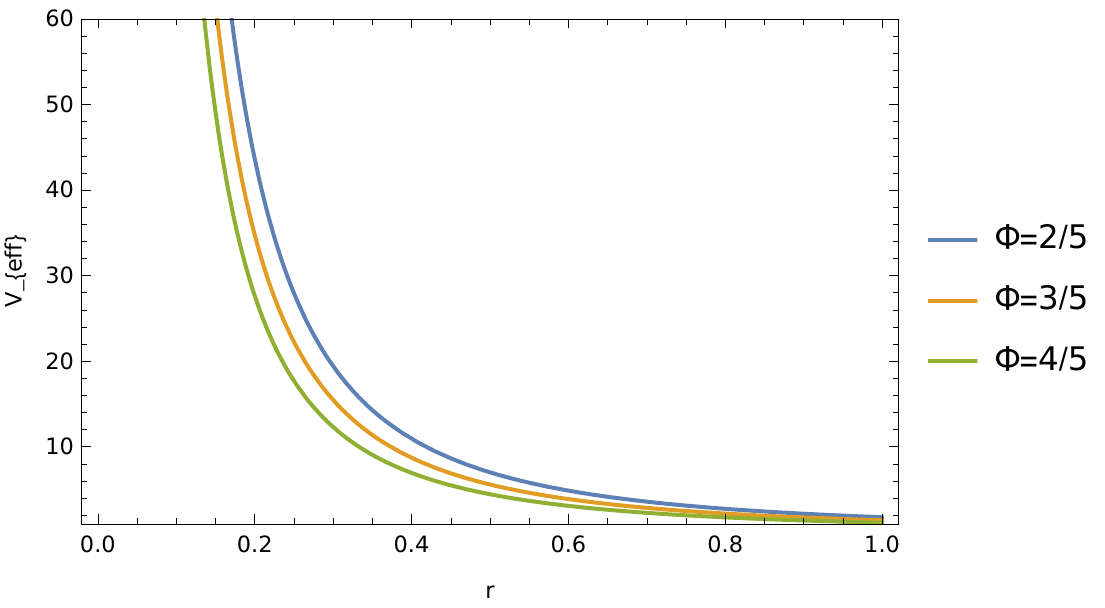}\quad
\includegraphics[width=3.0in,height=1.6in]{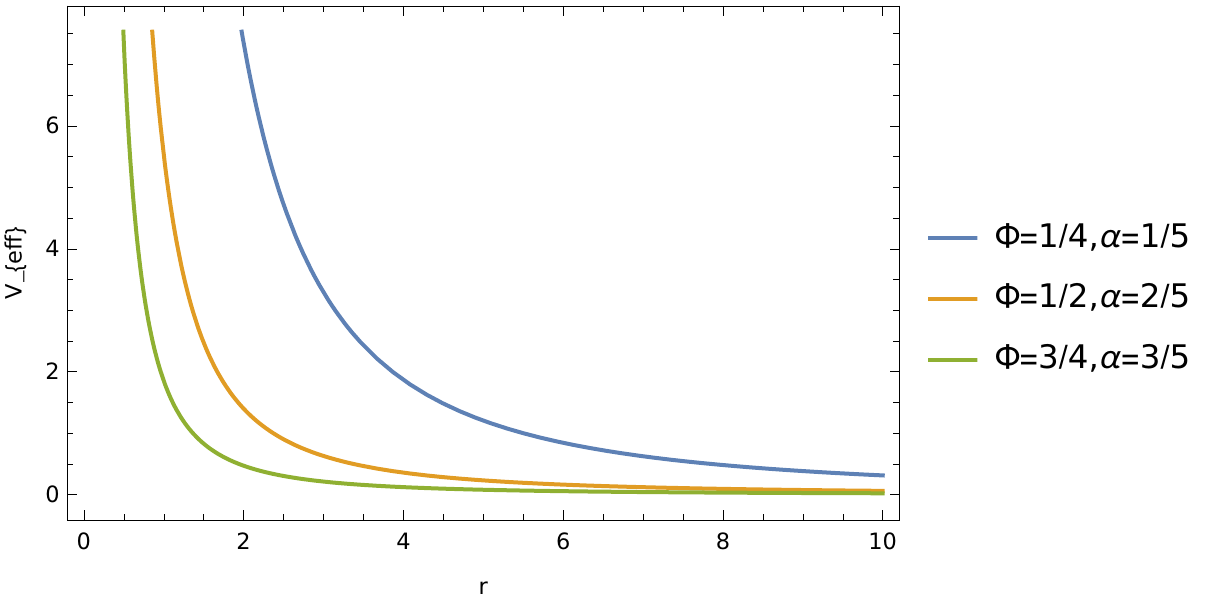}\quad
\caption{\small The effective potential $V_{eff}$ with radial distance $r$. Top left: with different values of the topological defect parameter $\alpha$ keeping fixed the magnetic flux $\Phi=1/2$; top right: with different values of the magnetic flux $\Phi$ keeping fixed the topological parameter $\alpha=3/4$; and bottom: with different values of the topological defect parameter $\alpha$ and the magnetic flux $\Phi$. Here, we have set the parameters $l=1=M=V_1$, $V_2=1/2$, $\delta=0.01$ and the units of these parameters are chosen in a system of units where $c=1=\hbar=G$.}
\label{fig: 3}
\end{figure*}

\begin{figure}
\centering
\includegraphics[width=3.4in,height=1.7in]{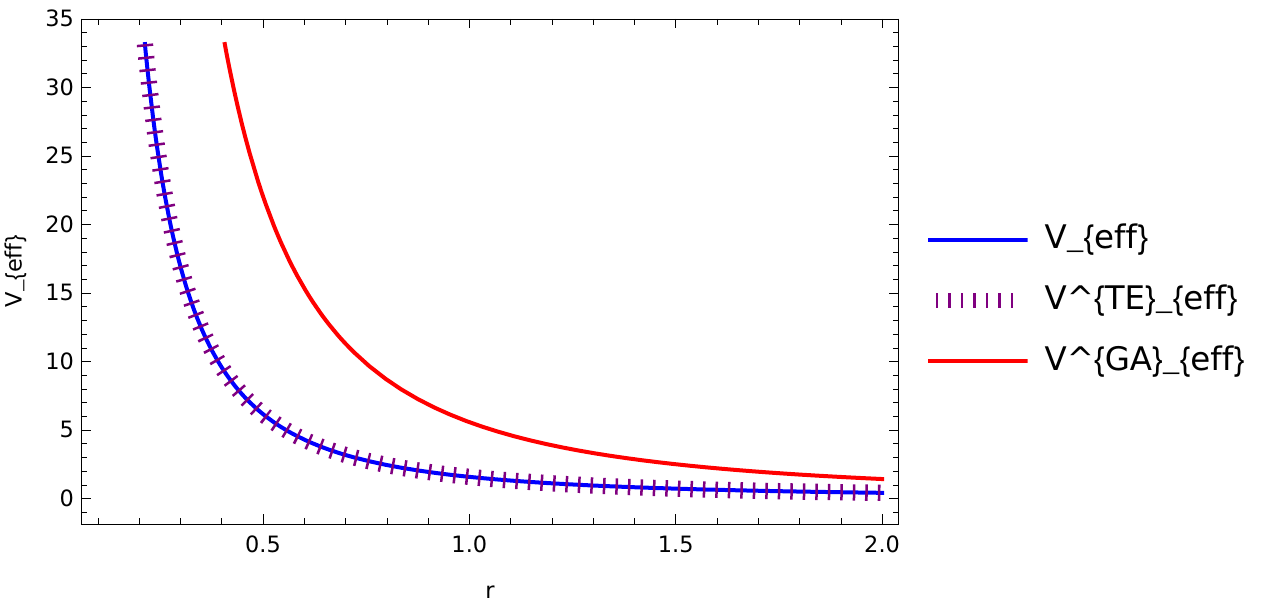}
\caption{\small Comparison of the effective potential of the system under different schemes. Red line for the Greene-Aldrich approximation, purple line for Taylor series expansion and blue line the original effective potential expression. Here, we have set the parameters $l=1=M=V_1=V_2$, $\delta=0.01$, $\alpha=1/2=\Phi$ and the units of these parameters are chosen in a system of units where $c=1=\hbar=G$.}
\label{fig: 4}
\end{figure}

With this potential, the radial equation (\ref{9}) can be written as
\begin{equation}
\psi''(r)+\Bigg[\gamma-\frac{j^2}{r^2}-\frac{2\,\eta}{r}\Bigg]\,\psi (r)=0,
\label{cc3}
\end{equation}
where 
\begin{eqnarray}
&&\gamma=\frac{2\,\mu\,(E+C)}{\alpha^2},\quad \eta=\frac{\mu\,B}{\alpha^2},\nonumber\\
&&j=\frac{1}{\alpha}\,\sqrt{(l-\Phi)\,(l-\Phi+1)+2\,\mu\,A}.
\label{cc4}
\end{eqnarray}

Transforming $\psi (r)=\sqrt{r}\,R(r)$ and then $x=2\,\sqrt{\gamma}\,r$ in Eq. (\ref{cc3}), we have
\begin{equation}
R'' (x)+\frac{1}{x}\,R'(x)+\Bigg(-\frac{\tau^2}{x^2}-\frac{1}{4}-\frac{\zeta}{x}\Bigg)\,R(x)=0,
\label{cc5}
\end{equation}
where $\tau=\sqrt{j^2+\frac{1}{4}}$ and $\zeta=\frac{\eta}{\sqrt{\gamma}}$.

The solution to the Eq. (\ref{cc5}) is given by
\begin{equation}
R(x)=x^{\tau}\,e^{-\frac{x}{2}}\,F(x),
\label{cc6}
\end{equation}
where $F(x)$ is an unknown function.

Substituting the solution (\ref{cc6}) in the Eq. (\ref{cc5}), we find the following equation
\begin{equation}
x\,F''(x)+(1+2\,\tau-x)\,F'(x)+\Big(-\tau-\zeta-\frac{1}{2}\Big)\,F(x)=0.
\label{cc7}
\end{equation}
Equation (\ref{cc7}) is the confluent hypergeometric differential equation form \cite{MAA,LJS}. The solution to this equation can be expressed in terms of a confluent hypergeometric function $F(x)=_1F_1 \Big(\tau+\zeta+\frac{1}{2},1+2\,\tau;x\Big)$ which is well-behaved for $x \to \infty$. For the bound-states solution, the function $_1F_1$ should be a finite degree polynomial of degree $n$, and the quantity
$\Big(\tau+\zeta+\frac{1}{2}\Big)=-n$, where $n = 0,1,2,....$. 

After simplification of the above condition, we have obtained the energy expression 
\begin{eqnarray}
&&E_{n,l}=-2\,\delta^2\,V_2\,\Bigg[1\nonumber\\
&&+\frac{\mu\,V_2}{\alpha^2\,\Big(n+\sqrt{\frac{(l-\Phi)\,(l-\Phi+1)+2\,\mu\,(V_1-V_2)}{\alpha^2}+\frac{1}{4}}+\frac{1}{2}\Big)^2}\Bigg].\quad\quad 
\label{cc8}
\end{eqnarray}
And the radial wave functions are given by
\begin{equation}
\psi(x)=\frac{1}{\sqrt{2}}\,\gamma^{-\frac{1}{4}}\,x^{\tau+\frac{1}{2}}\,e^{-\frac{x}{2}}\,_1F_1(-n,1+2\,\tau;x).
\label{cc9}
\end{equation}
Equation Eq (\ref{cc8})-(\ref{cc9}) is the eigenvalue solution of a non-relativistic particle under the influence of quantum flux field with inverse quadratic Yukawa plus inverse square potential in a point-like global monopole space-time background.

One can show that the obtained energy eigenvalue expression under different approximation schemes is different. We have generated a graph showing the comparison of the effective potential of the quantum system under the Greene-Aldrich scheme and Taylor series expansion (fig. 4). 

\section{Conclusions}

In this analysis, we investigated the eigenvalue solutions of the three-dimensional radial Schr\"{o}dinger equation under the influence of the Aharonov-Bohm flux field with a potential the superposition of an inverse quadratic Yukawa potential and inverse square potential in the background of topological defects produced by a point-like global monopole. We shown that the quantum motions of the particles depend on the non-trivial topological feature of the point-like global monopole and modified the result compared to flat space. Furthermore, the energy eigenvalue depends on the magnetic flux field and this dependence of the eigenvalue solution on the geometric quantum phase shows an analogue of the Aharonov-Bohm effect \cite{YA,MP} which is a quantum mechanical phenomena that has been studied by several researchers both in the non-relativistic and relativistic limit. we choose inverse  Since the chosen potential is an exponential-type, and thus we applied first the Greene-Aldrich approximation scheme in the centrifugal term that appeared in the radial equation and arrived a homogeneous differential equation after a few mathematical steps. This equation is then solved using the parametric Nikiforov-Uvarov method. The energy levels are given by the Eq. (\ref{21}) and the radial wave function by Eq. (\ref{22}). There, we studied the quantum system with an inverse quadratic Yukawa potential only ({\tt sub-section 2.2}) and solved the radial equation using the same NU-method. The energy levels are given by the Eq. (\ref{aa4}) and the radial wave functions by Eq. (\ref{aa5}). 

In {\tt section 3}, we investigated the same problem using a power series expansion as the chosen potential is an exponential-type. Taking up to the second-order of the exponential term, we derived the radial equation of the wave equation. Finally we solved this radial equation analytically and obtained the energy levels given by the Eq (\ref{cc8}) and the radial wave functions by Eq (\ref{cc9}). 

We have verified that the global effects of the space-time geometry characterized by the parameter $\alpha$ influences the energy eigenvalues and the radial wave functions of the non-relativistic particles and get them modified compared to the flat space result, that is, for $\alpha \to 1$. In addition, there is a influence of the quantum flux field on the eigenvalue solutions, and the dependence of the energy levels on the geometric quantum phase arises a persistent current that has applications in solid state physics which we will discuss in the future work. Also, one can study the thermodynamic properties of the quantum system by calculating the partition function using the formula $Z(\beta)=\sum\,e^{-\beta\,E_{n,l}}$, and then other properties, such as the Helmholtz free energy, mean free energy, specific heat, and entropy.

\section*{Conflict of Interest}

There are no conflicts of interests in this paper.

\section*{Data Availability Statement}

No data are generated or analysed in this paper.

\section*{Appendix: The parametric Nikiforov-Uvarov (NU) method}\label{appA}

\setcounter{equation}{0}
\renewcommand{\theequation}{A.\arabic{equation}}

The Nikiforov-Uvarov method is a helpful technique to calculate the exact and approximate energy eigenvalue and the wave functions of the Schr\"{o}dinger-like equation and other second-order differential equations of physical interest. According to this method, the wave functions of a second-order differential equation \cite{AFN}
\begin{equation}
\frac{d^2 \psi (s)}{ds^2}+\frac{(\alpha_1-\alpha_2\,s)}{s\,(1-\alpha_3\,s)}\frac{d \psi (s)}{ds}+ \frac{(-\xi_1\,s^2+\xi_2\,s-\xi_3)}{s^2\,(1-\alpha_3\,s)^2}\psi (s)=0
\label{A.1}
\end{equation}
are given by 
\begin{equation}
\psi (s)=s^{\alpha_{12}}(1-\alpha_3 s)^{-\alpha_{12}-\frac{\alpha_{13}}{\alpha_3}}P^{(\alpha_{10}-1,\frac{\alpha_{11}}{\alpha_3}-\alpha_{10}-1)}_{n}(1-2\alpha_3 s).
\label{A.2}
\end{equation}
And that the energy eigenvalues equation
\begin{eqnarray}
&&\alpha_2\,n-(2\,n+1)\,\alpha_5+(2\,n+1)\,(\sqrt{\alpha_9}+\alpha_3\,\sqrt{\alpha_8})\nonumber\\
&&+n\,(n-1)\,\alpha_3+\alpha_7+2\,\alpha_3\,\alpha_8+2\,\sqrt{\alpha_8\,\alpha_9}=0.\quad\quad
\label{A.3}
\end{eqnarray}
The parameters $\alpha_4,\ldots,\alpha_{13}$ are obatined from the six parameters $\alpha_1,\ldots,\alpha_3$ and $\xi_1,\ldots,\xi_3$ as follows:
\begin{eqnarray}
&&\alpha_4=\frac{1}{2}\,(1-\alpha_1),\quad \alpha_5=\frac{1}{2}\,(\alpha_2-2\,\alpha_3),\nonumber\\
&&\alpha_6=\alpha^2_{5}+\xi_1,\quad \alpha_7=2\,\alpha_4\,\alpha_{5}-\xi_2,\nonumber\\
&&\alpha_8=\alpha^2_{4}+\xi_3,\quad \alpha_9=\alpha_6+\alpha_3\,\alpha_7+\alpha^{2}_3\,\alpha_8,\nonumber\\ &&\alpha_{10}=\alpha_1+2\,\alpha_4+2\,\sqrt{\alpha_8},\nonumber\\
&&\alpha_{11}=\alpha_2-2\,\alpha_5+2\,(\sqrt{\alpha_9}+\alpha_3\,\sqrt{\alpha_8}),\nonumber\\
&&\alpha_{12}=\alpha_4+\sqrt{\alpha_8},\nonumber\\
&&\alpha_{13}=\alpha_5-(\sqrt{\alpha_9}+\alpha_3\,\sqrt{\alpha_8}).
\label{A.4}
\end{eqnarray}

A special case where $\alpha_3=0$, we find
\begin{equation}
\lim_{\alpha_3\rightarrow 0} P^{(\alpha_{10}-1,\frac{\alpha_{11}}{\alpha_3}-\alpha_{10}-1)}_{n}\,(1-2\,\alpha_3\,s)=L^{\alpha_{10}-1}_{n} (\alpha_{11}\,s),
\label{A.5}
\end{equation}
and 
\begin{equation}
\lim_{\alpha_3\rightarrow 0} (1-\alpha_3\,s)^{-\alpha_{12}-\frac{\alpha_{13}}{\alpha_3}}=e^{\alpha_{13}\,s}.
\label{A.6}
\end{equation}
Therefore the wave-function from (\ref{A.2}) becomes
\begin{equation}
\psi (s)=s^{\alpha_{12}}\,e^{\alpha_{13}\,s}\,L^{\alpha_{10}-1}_{n} (\alpha_{11}\,s),
\label{A.7}
\end{equation}
where $L^{(\beta)}_{n} (x)$ denotes the generalized Laguerre polynomial. 

The energy eigenvalues equation reduces to 
\begin{equation}
n\,\alpha_2-(2\,n+1)\,\alpha_5+(2\,n+1)\,\sqrt{\alpha_9}+\alpha_7+2\,\sqrt{\alpha_8\,\alpha_9}=0.
\label{A.8}
\end{equation}







\small
\begin{thebibliography}{99}


\bibitem{SMI} 
S. M. Ikhdair and R. Sever, J. Molec. Struct.: THEOCHEM {\bf 855}, 13 (2008).

\bibitem{SMI2}  
S. M. Ikhdair, Cent. Eur. J. Phys. {\bf 10}, 361 (2012).

\bibitem{SMI3} 
S. M. Ikhdair and M. Hamzavi, Few-Body Syst. {\bf 53}, 487 (2012).

\bibitem{CB} 
C. Berkdemir, J. Math. Chem. {\bf 46}, 139 (2009).

\bibitem{CB2} 
C. Berkdemir and Y. F. Cheng, Phys. Scr. {\bf 79}, 034003 (2009).

\bibitem{SMI4} 
S. M. Ikhdair and R. Sever, Int. J. Mod. Phys. {\bf C 19}, 221 (2008).

\bibitem{SMI5} 
S. M. Ikhdair and R. Sever, Int. J. Mod. Phys. {\bf A 21}, 6699 (2006).

\bibitem{SMI6} 
S. M. Ikhdair and R. Sever, Int. J. Mod. Phys. {\bf A 21}, 3989 (2006).

\bibitem{HH2} 
H. Hassanabadi, H. Rahimov, S. Zarrinkamar, Adv. High Energy Phys. {\bf 2011}, 458087 (2011).

\bibitem{SHD} 
S. H. Dong, Int. J. Quant. Chem. {\bf 109}, 701 (2009).

\bibitem{BJF} 
B. J. Falaye, Few-Body Syst. {\bf 53}, 557 (2012).

\bibitem{WG} 
W. Greiner and B. Muller, {\tt Quantum Mechanics: An Introduction}, Springer, Berlin, Germany (1994).

\bibitem{LDL} 
L. D. Landau and E. M. Lifshitz, {\tt Quantum Mechanics, the Non-relativistic Theory}, Pergamon, Oxford (1977). 

\bibitem{HH} 
H. Haken and H. C. Wolf, {\tt Molecular Physics and Elements of Quantum Chemistry: Introduction to Experiments and Theory}, Springer-Verlag, Berlin, Germany (1995).

\bibitem{FC} 
F. Constantinescu and E. Magyari, {\tt Probelms in Quantum Mechanics}, Pergamon Press, Oxford (1971).

\bibitem{DJG} 
D. J. Griffiths, {\tt Introduction to Quantum Mechanics}, New-Jersey: Prentice-Hall (2004).

\bibitem{CF} 
C. Furtado and F. Moraes, J. Phys. A: Math. Gen. {\bf 33}, 5513 (2000).

\bibitem{RV}  
R. L. L. Vitoria and H. Belich, Phys. Scr. {\bf 94}, 125301 (2019).

\bibitem{ALCO2} 
A. L. C. de Oliveira and E. R. B. de Mello, Int. J. Mod. Phys. {\bf A 18}, 3175 (2003).

\bibitem{ERBM33} 
E. R. B. de Mello and C. Furtado, Phys. Rev. {\bf D 56}, 1345 (1997).

\bibitem{ERBM44} 
A. L. Cavalcanti de Oliveira and E. R. B. de Mello, Int. J. Mod. Phys. {\bf A 18}, 2051 (2003).

\bibitem{ALCO} 
A. L. Cavalcanti de Oliveira and E R Bezerra de Mello, Class. Quantum Grav. {\bf 23}, 5249 (2006).

\bibitem{VBB} 
Geusa de A. Marques and Valdir B. Bezerra, Class. Quantum Gravit. {\bf 19}, 985 (2002).

\bibitem{PN} 
P. Nwabuzor, C. Edet, A. N. Ikot, U. Okorie, M. Ramantswana, R. Horchani, A. H. A.-Aty, G. Rampho, Entropy {\bf 23}(8), 1060 (2021).

\bibitem{MPP} 
F. Ahmed, Mol. Phys. {\bf 120}, e2124935 (2022). 

\bibitem{MPP2} 
F. Ahmed, Proc. R. Soc A {\bf 479}, 20220624 (2023), (arXiv: 2209.13490 [quant-ph]).

\bibitem{MPP3} 
F. Ahmed, Commun. Theor. Phys. (2023). https://doi.org/10.1088/1572-9494/acccdc (arXiv: 2210.04617 [hep-th]).

\bibitem{MPP4} 
F. Ahmed, Phys. Scr. {\bf 98}, 015403 (2023).

\bibitem{MPP5} 
F. Ahmed, EPL {\bf 141}, 25003 (2003).

\bibitem{MPP6} 
F. Ahmed, Mol. Phys. {\bf 121}, e2155596 (2023). 

\bibitem{MPP7} 
F. Ahmed, Int. J. Geom. Meths. Mod. Phys. {\bf 20}, 2350060 (2023). 

\bibitem{MPP8} 
F. Ahmed, Indian J. Phys. (2023); https://doi.org/10.1007/s12648-023-02590-6.

\bibitem{WCFS} 
W. C. F. da Silva and K. Bakke, Eur. Phys. J. C {\bf 79}, 559 (2019).

\bibitem{WCFS2} 
W. C. F. da Silva, K. Bakke and R. L. L. Vitoria, Eur. Phys. J. C {\bf 79}, 657 (2019).

\bibitem{ABHA} 
A. Boumali and H. Aounallah, Adv. High Energy Phys. {\bf 2018}, 1031763 (2018). 

\bibitem{MH2} 
M. Hamzavi, S. M. Ikhdair and B. I. Ita, Phys. Scr. {\bf 85}, 045009 (2012).

\bibitem{CAO} 
C. A. Onate, Afr. Rev. Phys. {\bf 8}, 0046 (2013).

\bibitem{BII2} 
B. I. Ita and A. I. Ikeuba, J. Atomic Mol. Phys. {\bf 2013}, 582610 (2013).

\bibitem{RLG} 
R L Greene and C Aldrich, Phys. Rev. {\bf A 14}, 2363 (1976).

\bibitem{MRS} 
M. R. Setare and S. Haidari, Phys. Scr. {\bf 18}, 065201 (2010)

\bibitem{WCQ} 
W. C. Qiang, K. Li and W. L. Chen, J. Phys. A: Math. Theor. {\bf 42}, 205306 (2009).

\bibitem{AFN} 
A. F. Nikiforov and V. B. Uvarov, Special Function of Mathematical Physics, Birkhauser, Basel (1988).

\bibitem{SZ} 
M. de Montigny, H. Hassanabadi, J. Pinfold and S. Zare, Eur. Phys. J. Plus {\bf 136}, 788 (2021).

\bibitem{SZ2} 
M. de Montigny, J. Pinfold, S. Zare and H. Hassanabadi, Eur. Phys. J. Plus {\bf 137}, 54 (2022).

\bibitem{YA} 
Y. Aharonov and D. Bohm, Phys. Rev. {\bf 115}, 485 (1959).

\bibitem{MP} 
M. Peshkin and A. Tonomura, {\tt The Aharonov-Bohm Effect}, Lecture Notes Phys. {\bf Vol. 340}, Springer-Verlag, Berlin, Germany (1989).

\bibitem{MAA} 
M. Abramowitz and I. Stegun, {\tt Handbook of Mathematical Functions}, {\bf Vol. 52}, Dover, New York (1965).

\bibitem{LJS} 
L. J. Slater, {\tt Confluent Hypergeometric Functions}, Cambridge University Press, Cambridge (1960).

\end{thebibliography}
\end{document}